\documentclass[12pt]{msml2021} %

\usepackage{bm,color,mathrsfs,verbatim,epstopdf,dcolumn,dsfont,cancel}

\usepackage{algorithm}
\usepackage{algorithmic}

\usepackage{soul}
\usepackage{mathtools}
\usepackage{booktabs}

\usepackage{color,xcolor}

\definecolor{mydarkblue}{rgb}{0,0.08,0.45}
\usepackage{flushend}

\usepackage{hyperref}
\hypersetup{
	colorlinks = true,
	citecolor=mydarkblue,
	urlcolor=mydarkblue
}

\DeclareMathOperator*{\argmin}{arg\,min}

\usepackage{mathtools}

\graphicspath{{new_figs/}, {figs/},{./}}

\newcommand{\agentname}{{RL-QAOA}}
\newcommand{\params}{{\bm\theta}}

\usepackage{amsmath,amssymb}

\usepackage{physics}

\renewcommand{\cite}{\citep}

\newcommand{\ad}{a^\mathrm{d}}
\newcommand{\Ad}{\mathcal A^{\mathrm d}}
\newcommand{\ac}{a^\mathrm{c}}
\newcommand{\Ac}{\mathcal A^{\mathrm c}}
\newcommand{\pc}{\pi_\params^\mathrm{c}}
\newcommand{\pd}{\pi_\params^\mathrm{d}}
\newcommand{\tc}{\tau^\mathrm{c}}
\newcommand{\td}{\tau^\mathrm{d}}

\newcommand{\pto}[1]{\pi_{\params_{\mathrm{old}}}^\mathrm{#1}}

\newcommand{\epc}{\epsilon^\mathrm{c}}
\newcommand{\epd}{\epsilon^\mathrm{d}}

\newcommand{\SC}{\mathcal S^{\mathrm{c}}}
\newcommand{\SD}{\mathcal S^{\mathrm{d}}}

\newcommand{\E}{\mathop{\mathbb{E}}}

\newcommand{\R}{\mathbb{R}}

\usepackage{indentfirst}
\usepackage{physics}
\usepackage{mathtools}

\newcommand{\eq}{\!=\!}
\newcommand{\plus}{\!+\!}
\newcommand{\minus}{\!-\!}

\newcommand{\logit}{\operatorname{logit}}

\usepackage{pifont}%
\newcommand{\cmark}{\ding{51}}%
\newcommand{\xmark}{\ding{55}}%

\newcommand{\nograd}{\multirow{2}{*}{{$\nabla$-free}}}

\usepackage[export]{adjustbox}

\usepackage{multirow}
\usepackage{colortbl}

\newcommand{\midsepremove}{\aboverulesep = 0mm \belowrulesep = 0mm \extrarowheight = 0.85ex}
\midsepremove
\newcommand{\midsepdefault}{\aboverulesep = 0.605mm \belowrulesep = 0.984mm \extrarowheight = 0mm}
\midsepdefault

\usepackage[pages=some,scale=1,angle=0,opacity=0.7]{background}

\usepackage{pgfplots}
\pgfplotsset{compat=1.12}

\usepackage{xcolor}

\definecolor{tblue}{HTML}{1f77b4}
\definecolor{torange}{HTML}{ff7f0e}
\definecolor{tgreen}{HTML}{2ca02c}
\definecolor{tred}{HTML}{d62728}

\definecolor{sblue}{HTML}{4c72b0}
\definecolor{sorange}{HTML}{dd8452}
\definecolor{sgreen}{HTML}{55a868}
\definecolor{sred}{HTML}{c44e52}

\usepackage{enumitem,amssymb}
\newlist{todolist}{itemize}{2}
\setlist[todolist]{label=$\square$}
\usepackage{pifont}

\usepackage[scientific-notation=true]{siunitx}

\newcommand*\samethanks[1][\value{footnote}]{\footnotemark[#1]}

\title[\agentname: Reinforcement learning for quantum control using autoregressive policy]{Noise-Robust End-to-End Quantum Control \\
using Deep Autoregressive Policy Networks}

\usepackage{times}
\msmlauthor{%
 \Name{Jiahao Yao} \Email{jiahaoyao@berkeley.edu}\\
 \addr Department of Mathematics, University of California, Berkeley\\ Berkeley,  CA 94720, USA
  \AND
  \Name{Paul K\"{o}ttering}\thanks{ P.K. \& H.G. contributed equally and listed
 by a coin flip} \\
 \addr Department of Mathematics, University of California, Berkeley\\ Berkeley,  CA 94720, USA
  \AND
  \Name{Hans Gundlach}\samethanks[1]\\
 \addr Department of Mathematics, University of California, Berkeley\\ Department of Physics, University of California, Berkeley\\ Berkeley,  CA 94720, USA
 \AND
  \Name{Lin Lin}  \\
 \addr Department of Mathematics, University of California, Berkeley\\
Computational Research Division, Lawrence Berkeley National Laboratory\\
Challenge Institute for Quantum Computation, University of California, Berkeley\\
Berkeley, CA 94720, USA
 \AND
 \Name{Marin Bukov} \Email{mgbukov@phys.uni-sofia.bg}\\
 \addr Department of Physics, St.~Kliment Ohridski University of Sofia\\ 5 James Bourchier Blvd, 1164 Sofia, Bulgaria%
}

\begin{document}

\maketitle

\begin{abstract}

    Variational quantum eigensolvers have recently received increased attention, as they enable the use of quantum computing devices to find solutions to complex problems, such as the ground energy and ground state of  strongly-correlated quantum many-body systems. In many applications, it is the optimization of both continuous and discrete parameters that poses a formidable challenge.
    Using reinforcement learning (RL), we present a hybrid policy gradient algorithm capable of simultaneously optimizing continuous and discrete degrees of freedom in an uncertainty-resilient way. The hybrid policy is modeled by a deep autoregressive neural network to capture causality.
    We employ the algorithm to prepare the ground state of the nonintegrable quantum Ising model in a unitary process, parametrized by a generalized quantum approximate optimization ansatz: the RL agent solves the discrete combinatorial problem of constructing the optimal sequences of unitaries out of a predefined set and, at the same time, it optimizes the continuous durations for which these unitaries are applied. We demonstrate the noise-robust features of the agent by considering three sources of uncertainty: classical and quantum measurement noise, and errors in the control unitary durations.
    Our work exhibits the beneficial synergy between reinforcement learning and quantum control.

\end{abstract}

\section{Introduction} \label{sec:intro}

The last decade has seen impressive breakthroughs in Machine Learning (ML), ranging from image classification~\cite{lecun2015deep, krizhevsky2017imagenet} to mastering complex video and board games~\cite{mnih2013playing, silver2016mastering}. ML algorithms have opened the door to solving major scientific challenges, hitherto considered intractable, such as protein modelling~\cite{rao2019evaluating} and folding~\cite{john2020high}, or molecular dynamics simulations~\cite{lu202086}.

Deep learning tools and methods quickly found their way into the field of physics~\cite{dunjko2018machine,mehta2019high,carleo2019machine,carrasquilla2020machine}:
Supervised learning was found efficient in identifying phase transitions and analyzing experimental data~\cite{carrasquilla2017machine,vanNieuwenburg2017learning,bohrdt2019classifying,rem2019identifying}.
Unsupervised learning brought a new class of variational many-body wavefunctions~\cite{carleo2017solving}, as well as methods to perform tomography on many-body quantum states~\cite{torlai2018neural}, find conservation laws from data~\cite{iten2020discovering}, identify phase transitions~\cite{wang2016discoverning,kottmann2020unsupervised}, Hamiltonian learning~\cite{valenti2019hamiltonian}, etc.
Reinforcement learning (RL)~\cite{sutton2018reinforcement} brought strategies for navigating turbulent flows~\cite{reddy2016learning,colabrese2017flow,Bellemare2020},
and even exploring the string landscape~\cite{halverson2019branes}.

The variational character of ML models combined with their intrinsic optimization procedure, provide a natural playground for applications in quantum control~\cite{schafer2020differentiable,wang2020bayesian,sauvage2019optimal,fosel2020efficient,nautrup2019optimizing,albarran2018measurement,sim2020adaptive,wu2020end,wu2020active, abhinav2020natural}.
Due to the close relationship between control theory and reinforcement learning, the control of quantum systems has become a major application area of RL algorithms in physics. Notable examples include policy gradient \cite{niu2019universal, fosel2018reinforcement, august2018taking, porotti2019coherent,wauters2020reinforcement,yao2020policy,sung2020towards}, Q-learning \cite{chen2013fidelity, bukov2018reinforcement, bukov2018day, sordal2019deep, bolens2020reinforcement} and AlphaZero \cite{dalgaard2020global}.

Over the years, the physics community has also developed a number of successful quantum control algorithms~\cite{khaneja2005optimal,caneva2011chopped, peruzzo2014variational, dalgaard2020hessian, magann2020digital, magann2020pulses}, including GRAPE, CRAB, and VQE. A prominent example of the latter is Quantum Approximate Optimization Algorithm (QAOA)~\cite{farhi2014quantum}, whose versatility allows for solving complex combinatorial problems using quantum computers~\cite{garcia2019quantum, dong2019robust,khairy2019reinforcement,khairy2020learning,yao2020reinforcement,tabi2020quantum,bravyi2020hybrid}.
Quantum control algorithms, such as CRAB or QAOA, come with an ingenious physics-informed variational ansatz for the structure of control protocols. RL algorithms, on the other hand, are model-free and resilient to uncertainty.
Hence, a natural question emerges as to how one can combine the benefits offered by RL and quantum control in a unified framework.

In this paper, our aim is to deploy a generalized QAOA ansatz in combination with an end-to-end deep RL algorithm for a versatile continuous-discrete quantum control [Sec.~\ref{sec:prelim}].
We adopt the continuous degrees of freedom of QAOA which offer an increased control accuracy. Additionally, we consider an enhanced variational control ansatz which contains a larger space to select the building blocks of the protocols from; this introduces a second, discrete combinatorial optimization problem.
The resulting algorithm, \agentname{}, realizes greater gains by striking a balance between robustness and versatility:
it is resilient to various kinds of uncertainty, a property shared with PG-QAOA~\cite{yao2020policy}; at the same time, \agentname{} has access to the more general variational counter-diabatic (CD) driving ansatz~\cite{demirplak_05,masuda_09,guery2019shortcuts} through CD-QAOA~\cite{yao2020reinforcement}.

However, \agentname{} presents a number of new challenges, cf.~Sec.~\ref{sec:algo}.
It requires a mixed continuous-discrete action space so that the RL agent can
construct a control protocol by
optimizing the order in which unitaries appear in the control sequence; simultaneously, the agent has to also choose the continuous duration to apply each unitary. This requires the use of a suitable ML model to approximate the policy, which allows us to build in temporal causality.
Therefore, an essential building block of \agentname{} is a novel monolithic deep autoregressive policy network that handles continuous and discrete actions on equal footing.
To train our RL agent, we derive an extension of Proximal Policy Optimization (PPO)~\cite{schulman2017proximal} to hybrid discrete-continuous policies.

We apply \agentname{} to find the ground state of a nonintegrable chain of interacting spin-$1/2$ particles (a.k.a.~qubits) in a fixed amount of time,
cf.~Sec.~\ref{sec:Ising-1/2}.
The mixed discrete-continuous degrees of freedom allow the RL agent to construct a short protocol sequence away from the adiabatic regime.
We test the agent's behavior in a strongly stochastic environment, by considering three different kinds of noise:
classical and quantum measurement noise, and errors in the control unitary gate duration.
In Sec.~\ref{sec:experiment}, we demonstrate that \agentname{} is insensitive to the types of noise applied, and outperforms previously developed algorithms based on QAOA in the regime of strong noise.

\section{Preliminaries}
\label{sec:prelim}

We start the discussion by introducing the QAOA ansatz used in quantum control. Following a short overview of reinforcement learning terminology, we review two RL-based QAOA algorithms --- PG-QAOA and CD-QAOA --- which we aim to blend into a homogeneous hybrid in Sec.~\ref{sec:algo}. The resulting new algorithm combines the benefits of the generalized variational QAOA ansatz, with an RL algorithm performing both continuous and discrete control simultaneously.

\subsection{QAOA for Ground State Preparation} \label{sec:qc-basics}

Of particular interest in the quest for designing new materials with novel features (such as room-temperature superconductors, or topological quantum computers), is the study of ground state properties in quantum many-body physics. Quantum simulators provide an ideal platform to bring together both theory and experiment; yet, they require the ability to prepare a system in its ground state -- a formidable challenge for modern quantum computing devices, due to the presence of various sources of uncertainty and noise. The Quantum Approximate Optimization Algorithm (QAOA)~\cite{farhi2014quantum} provides a widely used state-of-the-art ansatz for this purpose.

Consider a quantum system of $N$ qubits, described by the Hamiltonian $H$. Starting from an initial quantum state $\ket{\psi_i}$, in QAOA we apply two alternating unitary evolution operators (i.e.~quantum gates) \citep{farhi2014quantum}:
\begin{equation}
    \ket{\psi(T)}=U(\{\alpha_j, \beta_j\}_{j=1}^p) \ket{\psi_i}= e^{-i H_2 \beta_p} e^{-i H_1 \alpha_p} \cdots e^{-i H_2 \beta_1} e^{-i H_1 \alpha_1}\ket{\psi_i}.
    \label{eqn:qaoawf}
\end{equation}
The dynamics are generated by the time-independent operators $H_1$ and $H_2$, applied for a duration of $\alpha_j\geq0$ and $\beta_j\geq0$, respectively ($j=1,2,\cdots, p$ with $p\in\mathbb{N}$). We refer to $q \eq 2p$ as the total circuit depth.
In order to apply QAOA to many-body systems~\cite{ho2019efficient}, the protocol durations $\{(\alpha_j,\beta_j)\}_{j=1}^{p}$ are variationally optimized to minimize the expected value of the energy density $\mathcal{E}(\{\alpha_j, \beta_j\}_{j=1}^p) \eq N^{-1}  { \mel{\psi(T)} H { \psi(T)}}$
:
\begin{equation}
    \{\alpha_j^\ast, \beta_j^\ast\}_{j=1}^p \eq \argmin_{\{\alpha_j, \beta_j\}_{j=1}^p} \mathcal{E}(\{\alpha_j, \beta_j\}_{j=1}^p),\quad
    \sum_{j=1}^p (\alpha_j+\beta_j)=T
    .
    \label{eqn:qaoa}
\end{equation}
The additional constraint $\sum_{j=1}^p (\alpha_j+\beta_j)=T$ is required for the resulting protocol to remain in the regime of practical applications, and also for a fair comparison between different algorithms.

As a concrete example to keep in mind, consider the spin-$1/2$ Ising Hamiltonian
\begin{equation}
    \label{eq:IM}
    H \!=\!H_1\!+\!H_2,\qquad
    H_1\!\!=\!\! \sum_{i=1}^N J S^z_{i+1}S^z_i \!+\! h_z S^z_i,\quad
    H_2= \sum_{i=1}^N h_xS^x_i,
\end{equation}
where $[S^\alpha_k,S^\beta_j]=i\delta_{kj}\varepsilon^{\alpha\beta\gamma}S^\gamma_j$ are the spin-$1/2$ operators.
We are interested in preparing the ground state of $H$, starting from a spin-up polarized initial product state. More details about the physical system are discussed later on in Sec.~\ref{sec:Ising-1/2} and App.~\ref{app:durations}.

\subsection{Reinforcement Learning (RL)}
\label{sec:rl-basics}

While QAOA defines a variational ansatz to prepare ground states in a unitary process, it does not yet provide a self-contained optimization procedure to find the optimal protocol durations. A universal optimization framework is presented by RL~\cite{sutton2018reinforcement}.

Reinforcement learning comprises a powerful set of algorithms designed to solve control problems. In RL, an agent aims to find a policy $\pi$ which solves a specific task in a trial-and-error approach based on interactions with the agent's environment. Consider a finite-horizon Markov Decision Process (MDP) defined by the tuple $(\mathcal S, \mathcal A,p,r)$ where $\mathcal S$ and $\mathcal A$ are the state and action spaces, respectively, and $p: \mathcal S\times \mathcal S \times \mathcal A \rightarrow [0,1]$ defines the transition probability which governs the environment dynamics. Upon selecting an action $a\in\mathcal{A}$, the environment transitions to a new state $s\to s'\in\mathcal{S}$, and emits a reward $r: S\times A \rightarrow \mathbb{R}$, which the RL agent uses to select subsequent actions. The action $a_j\in\mathcal{A}$ to be selected in a given state $s\in\mathcal{S}$ is determined probabilistically by the instantaneous policy $\pi(a_j|s_j): \mathcal A \times  \mathcal S \rightarrow [0, 1]$. For a given policy $\pi$, this process generates a trajectory $\tau=(s_1, a_1, ...., a_{q}, s_{q+1})$
with probability $\tau \sim \mathbb P^{\pi}( \cdot )$. Here, $\mathbb{P}^{\pi}( \tau ) = p_0(s_1)\pi(a_1|s_1)p(s_2|s_1, a_1)\cdots \pi(a_{q}| s_q)p(s_{q+1}|s_{q}, a_{q})$, the episode/trajectory length is $q$, and $p_0$ is the initial state distribution.
The objective in RL is to find the optimal policy, i.e.~the policy which maximizes the total expected return: $\mathbb{E}_{\tau \sim \mathbb P^{\pi}}\left[\sum_{j=1}^{q}  r(s_j,a_j)\right].$

\subsection{Policy Gradient Quantum Approximate Optimization Algorithm (PG-QAOA)} \label{sec:PG-QAOA}

A reinforcement learning based approach to QAOA was recently introduced in Ref.~\cite{yao2020policy}, using a policy gradient algorithm. The basic idea behind PG-QAOA is to let the RL agent select the durations $\{\alpha_j,\beta_j\}$, which constitute a continuous action space $\Ac$.
However, casting the quantum control problem within the RL framework comes with certain challenges.
The first challenge is that quantum states cannot be directly measured in experiments, which poses questions about the proper definition of the RL state space. To remedy this in an environment following deterministic Schr\"odinger dynamics, it was suggested to fix the initial quantum state, and define the RL state as the trajectory of actions $s_j=(\ac_1, \cdots, \ac_{j-1})=(\alpha_1, \beta_1, \cdots)$ up to episode step $j$~\cite{bukov2018reinforcement}; this definition is inferior to using the full state, but it allows to accommodate the experimental constraint, so we adopt it in this study as well. Alternatively, one could use the expectation values of observables to define an RL state~\cite{wauters2020reinforcement}.
The second challenge is the sparsity of the reward signal -- a quantum measurement is allowed only once at the end of each episode, since projective measurements collapse the quantum wavefunction and the quantum state is lost irreversibly.

Since the protocol durations are continuous degrees of freedom, we need an RL method for \textit{continuous} optimization.
PG-QAOA defines the simplest ansatz: $q\eq 2p$ independent Gaussian distributions to parameterize the policy, one for each duration $\{\alpha_j, \beta_j\}_{j=1}^p$ in Eq.~\eqref{eqn:qaoa}.
Since a Gaussian distribution is uniquely determined by its mean $\mu$ and standard deviation $\sigma$, we need a total of $2p$ independent variational parameters ${\bf\params} = \{\mu_{\alpha_j}, \sigma_{\alpha_j}, \mu_{\beta_j}, \sigma_{\beta_j}\}_{j=1}^p$ to parametrize the policy $\pi_{\params}$ as:
\begin{equation}
    \pi_{\params}(\{\alpha_j, \beta_j\}_{j=1}^p) = \prod_{j=1}^p \pi(\alpha_j ;\kappa_{\alpha_j}, \xi_{\alpha_j}) \pi(\beta_j ;\kappa_{\beta_j}, \xi_{\beta_j}),
    \label{eqn:prob}
\end{equation}
where $\kappa_{\alpha_j}=\mu_{\alpha_j}$, $\kappa_{\beta_j}=\mu_{\beta_j}$ are the means, and $\xi_{\alpha_j}=\sigma_{\alpha_j}$, $\xi_{\beta_j}=\sigma_{\beta_j}$ are the variances of the Gaussian policy.
The actual protocol durations are thus sampled according to $\alpha_j \sim \mathcal{N}(\mu_{\alpha_j},\sigma^2_{\alpha_j})$, and similarly for $\beta_j$. As was shown in Ref.~\cite{yao2020policy}, despite its simplicity, PG-QAOA defines a particularly noise-robust algorithm. In the presence of various kinds of noise, it readily outperforms a number of alternative gradient-free optimization algorithms.

\begin{figure*}[t!]
    \centering
    \begin{minipage}{.5\textwidth}
        \centering
        \includegraphics[width=1.0\textwidth]{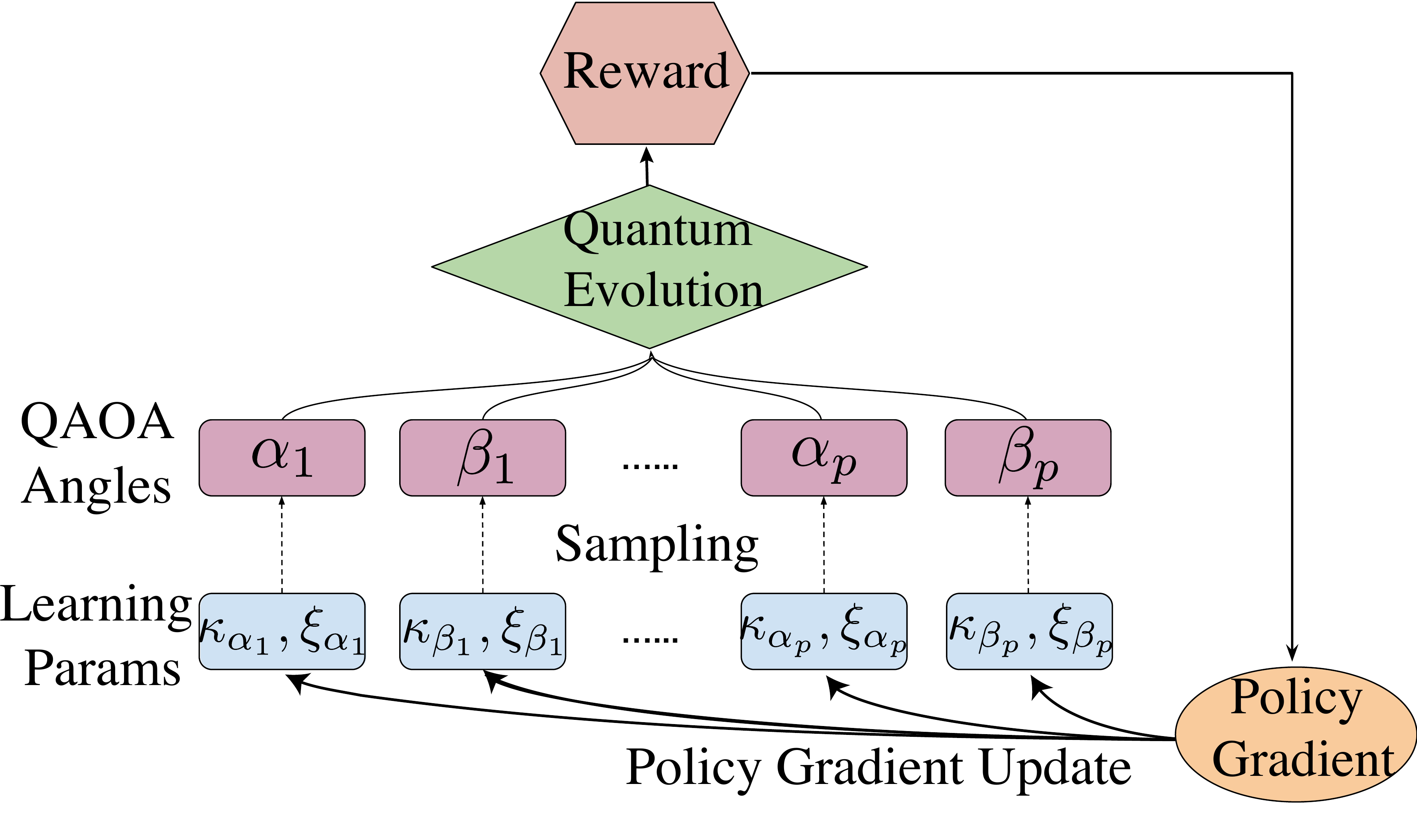} \\
        PG-QAOA
    \end{minipage}%
    \begin{minipage}{0.5\textwidth}
        \centering
        \includegraphics[width=1.0\textwidth]{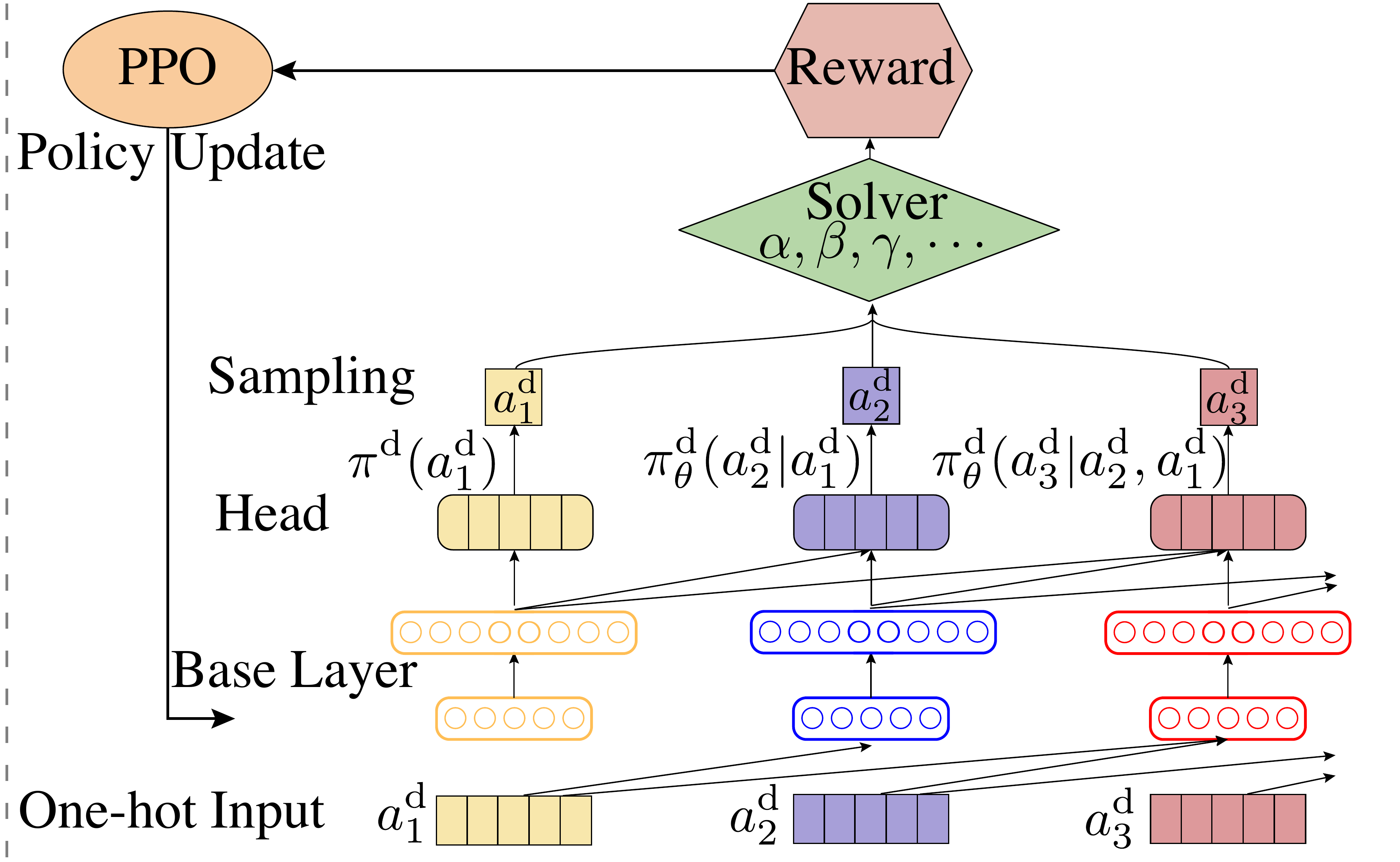} \\
        CD-QAOA
    \end{minipage}
    \caption{
    \label{fig:arpgqaoa-arch-diag}
    \small Schematic diagram for PG-QAOA [left, see Sec.~\ref{sec:PG-QAOA}] and CD-QAOA [right, see Sec.~\ref{sec:CD-QAOA}]. The PG-QAOA samples continuous QAOA-angles from its policy and variationally updates the policy parameters via policy gradient; CD-QAOA autoregressively samples the gate sequences for the generalized QAOA ansatz and employs the gradient-free solver (Powell algorithm) to solve for their corresponding durations. The policy network is updated via Proximal Policy Optimization~(PPO).   For a comparison with \agentname, cf.~Fig.~\ref{fig:arnet}.}

    \vspace{-2mm}
\end{figure*}

For this study, the original PG-QAOA implementation~\cite{yao2020policy} is not directly applicable, and a modification is required.
First, the extensive scaling with increasing the number of qubits suggests us to use as a cost function the energy density, rather than the many-body fidelity; in doing this, the algorithm no longer requires an explicit reference to the target ground state we are searching for.
Second, the original PG-QAOA algorithm does not support an easy implementation of the protocol duration constraint $ \sum_{j= 1}^p (\alpha_j + \beta_j) \eq T$. Here, in order to do a fair comparison among different algorithms, we enforce this constraint. Note that this is a non-trivial task for policy gradient, for three reasons: (i) protocol durations are sampled from a Gaussian distribution which has unbounded support, (ii) a Gaussian policy supports negative as well as positive samples (yet we require $\alpha_j,\beta_j\geq 0$ for a physical time duration), and (iii) sampled values, even if bounded and nonnegative, are always random, and hence one needs to additionally fix their total sum. We consider two different approaches to resolving (i) and (ii), and apply a normalization trick to fix (iii).

The first approach we consider is to define the policy using the Beta distribution~\cite{chou2017improving}, i.e.~$\alpha_j,\beta_j \sim \mathrm{B}(\kappa,\xi)$, instead of a Gaussian, and learn the two nonnegative parameters $\kappa,\xi$. Since the Beta distribution is defined on the interval $\Ac=[0,1]$, it solves the boundedness and positivity problems. The policy is given by Eq.~\eqref{eqn:prob} with $\pi(x ; \kappa, \xi)=\frac{\Gamma(\kappa+\xi)}{\Gamma(\kappa) \Gamma(\xi)} x^{\kappa-1}(1-x)^{\xi-1}$ the probability density for the Beta distribution; $\Gamma$ denotes the Gamma function. Note that the number of independent variational parameters ${\bf\params} = \{\kappa_{\alpha_j}, \xi_{\alpha_j}, \kappa_{\beta_j}, \xi_{\beta_j}\}_{j=1}^p$, remains equal to $4p$.

In the second approach, we pass the output of the Gaussian distribution through a sigmoid activation function~\cite{haarnoja2018soft}. Due to the boundedness of the sigmoid function, this restricts the range of all actions/durations to the nonnegative interval $\Ac=[0,1]$. Hence, our policy is given by Eq.~\eqref{eqn:prob}
where $\pi(x; \kappa, \xi) \eq \frac{1}{x (1\minus x)}\frac{1}{\sqrt{2\pi \xi^2}}\exp\left(- \frac{(\logit(x)-\kappa)^2 }{2\xi^2}\right)$ is the probability density for the Sigmoid Gaussian distribution $\mathcal{SN}(\kappa, \xi^2)$\footnote{$\mathcal{SN} (\kappa, \xi^2)$ is short-hard notation for Gaussian distribution
    $\mathcal N(\kappa, \xi^2)$ under the sigmoid transformation $f(x)$.}.
Here, the logit function, $\logit(x) \eq \log x - \log(1-x)$, is the inverse of the sigmoid function $f(x) = 1/(1+\exp(-x))$, and the factor $\frac{1}{x (1 - x)}$ is the inverse Jacobian of $x \eq f(y)$ over $y$ [cf.~App.~\ref{app:sigmoid-gaussian}]. Notice how, the action output of this policy is forced within the interval $[0,1]$ by construction, without changing the total number of independent variational parameters $\params$.

Finally, to fix the total protocol duration, (iii), we normalize the sum of durations manually according to
$\alpha_j \eq \frac{\alpha_j}{ \sum_{j\eq 1}^p (\alpha_j \plus \beta_j)} T,\; \beta_j \eq \frac{\beta_j}{ \sum_{j\eq 1}^p (\alpha_j \plus \beta_j)} T$.
We note that the normalization procedure is considered part of the RL environment, i.e.~no gradients are passed through it. In essence, it becomes part of the reward function. This requires us to slightly re-define the meaning of the policy: it generates the bare protocol durations before the normalization; to minimize energy, the durations need the extra normalization. We mention in passing that this is not the only way to hold the protocol duration $T$ fixed: alternatives include using the constraint to fix the last protocol duration $\beta_p$, or the addition of an extra penalty term to the cost function.

\subsection{Quantum Approximate Optimization Ansatz based on Counter-Diabatic Driving} \label{sec:CD-QAOA}

In conventional QAOA, there are two possible gates, corresponding to the two unitaries $U_j=\exp(-i\alpha_j H_j), j=1,2$. Therefore, there exist only two distinct sequences of unitaries: $\tau^\mathrm d_1 = U_1 U_2 U_1 U_2\cdots$ and $\tau^\mathrm d_2 = U_2 U_1 U_2 U_1\cdots$. A generalization of this ansatz was considered in Ref.~\cite{yao2020reinforcement}, where an RL agent was given the complex combinatorial task to construct the sequence of unitaries $\tau^\mathrm d$, out of a predefined set $\Ad$ of $|\Ad|$ gates/unitaries. As for the gate duration notation, we will use $\alpha_j$ for all durations instead of alternating $\alpha_j$, $\beta_j$ due to a general ansatz. This set can, in principle, be chosen arbitrarily; however, one can also make a more physics-informed choice, e.g., inspired by counter-diabatic driving in the case of quantum-many-body systems [cf.~Sec.~\ref{subsec:actions}].
In the latter case, the resulting generalized algorithm, called CD-QAOA, was demonstrated to drastically enhance the variational ansatz of QAOA when applied to many-body quantum chains, allowing for shorter circuit depths at no cost in performance~\cite{yao2020reinforcement}.

Similar to PG-QAOA, CD-QAOA does \emph{not} use the quantum wavefunction to perform the optimization, and the state is $s_j = (\ad_{1}, \cdots, \ad_{j-1})$ at episode step $j$. Rewards are given once per episode, in the end, and are defined by the (negative) energy density. However, the action space is given by the set of $|\Ad|$ unitary gates from which the protocol sequence $\td$ are selected; it does not involve the continuous protocol durations which are found as part of the RL environment; to do this, in this study, we use the gradient-free Powell algorithm~\cite{powell1964efficient} instead of the gradient-based SLSQP algorithm~\cite{kraft1988software} presented in the original CD-QAOA paper.

Apart from the low-level optimization mentioned above, CD-QAOA adopts a two-level optimization schedule~\cite{li2020quantum, melnikov2020setting}: high-level discrete optimization is used to construct the optimal sequence $\td$ out of the available set of unitaries. For this purpose, in Ref.~\cite{yao2020reinforcement}, it was suggested to employ Proximal Policy Optimization\cite{schulman2017proximal} (PPO), an advanced variant of policy gradient, aided by a deep autorgressive neural network to implement causality:
\begin{equation}
    \label{eq:autoreg_pi}
    \pd\left(\ad_{1}, \ad_{2}, \cdots, \ad_{q}\right)=\pd\left(\ad_{1}\right) \prod_{j=2}^{q} \pd\left(\ad_{j} \mid \ad_{1}, \cdots, \ad_{j-1}\right).
\end{equation}
Each factor in the product above is a categorical distribution over the action space.
We point out that the search for the optimal sequence $\tau^\mathrm d $ represents a \textit{discrete} optimization problem. This should be contrasted with the low-level continuous optimization employed by QAOA to find the optimal durations $\{\alpha_j\}_{j=1}^q$, carried out using the Powell solver. Since the Powell solver only deals with the bounded optimization, we apply the same normalization trick to enforce the total duration constraint.

Given the complete protocol sequence $\tau^\mathrm d =(\ad_{1}, \cdots, \ad_{q})$, we can construct the unitary process
\begin{equation}
    \label{eq:U_ansatz}
    U(\{\alpha_j\}_{j=1}^q ,\tau^\mathrm d)\!=\!\prod_{j=1}^q U_{\tau_j^\mathrm d }(\alpha_j)
\end{equation}
which we use as a generalized QAOA ansatz. The sequence $\tau^\mathrm{d}$, and the durations $\{\alpha_j\}_{j=1}^q$ are found by minimizing the energy density, cf.~Eq.~\eqref{eqn:qaoa}. In doing so, we impose an extra constraint that the same action cannot be taken twice in a row, for otherwise one can add the corresponding durations and optimize them together.

\subsubsection{Possible Choice of Unitaries based on Counter-Diabatic Driving} \label{subsec:actions}

In order to construct the unitary $U(\{\alpha_j\}_{j=1}^q ,\tau^\mathrm d ) = \prod_{j=1}^q \exp(-i\alpha_j H_{\td_j})$  from Eq.~\eqref{eq:U_ansatz}, the RL agent needs to select the sequence $\td$ of subprocess generators $\Ad$. Hence,
at every step $j$ in the RL episode, the agent's action consists of a choice of a Hermitian operator $H_{\td_j}\in\Ad$%

The set of discrete actions $\Ad$ consists of the available possible controls in an experiment. In Refs.~\cite{yao2020reinforcement, hegade2020shortcuts, ding2020breaking}, it was shown that a particularly suitable choice of actions for ground state preparation in quantum many-body systems, is given by terms appearing in the series of the variational adiabatic gauge potential, designed for many-body counter-diabatic driving~\cite{sels2017minimizing}. These terms provide shortcuts in the Hilbert space that may significantly decrease the time required to prepare the ground state. For brevity, below we just list the generator set $\Ad$ for the spin$-1/2$ Ising model, cf.~Eq.~\eqref{eq:IM}, which the RL agent has access to, and refer the interested readers to Ref.~\cite{yao2020reinforcement} for more details: $\Ad=\{H_1, H_2, Y, X|Y, Y|Z\}$, with $Y=\sum_i S^y_{i}$, $X|Y=\sum_i S^x_{i}S^y_{i+1} + S^y_{i}S^x_{i+1}$, and $Y|Z=\sum_i S^y_{i}S^z_{i+1} + S^z_{i}S^y_{i+1}$; $H_1, H_2$ are defined in Eq.~\eqref{eq:IM}.

We emphasize that this is just one particular choice for $\Ad$. In practice, the algorithm is agnostic to the discrete action space which is determined by the available controls for the system of interest: e.g., on a quantum computer, these can be the set of local gates, etc.

\begin{table}[t!]
    \centering
    \midsepremove
    \begin{tabular}{c||c|c|c|>{\columncolor[gray]{0.95}}c}
        \toprule
        Method            & QAOA                    & PG-QAOA                 & CD-QAOA                        & \cellcolor[gray]{0.9}\agentname \\
        \midrule \midrule
        protocol sequence & \multirow{2}{*}{\xmark} & \multirow{2}{*}{\xmark} & \multirow{2}{*}{$\nabla$-free} &                                 \\ optimization (discrete) & &&& \multirow{-2}{*}{$\nabla$-free} \\
        \midrule
        gate durations    & \nograd                 & \nograd                 & \nograd                        &                                 \\ optimization (continuous) &  &  &  & \multirow{-2}{*}{$\nabla$-free} \\
        \midrule
        RL optimization   & \xmark                  & continuous              & discrete                       & continuous \&  discrete         \\
        \midrule
        noise-robust      & \xmark                  & \cmark                  & \xmark                         & \cmark                          \\
        \midrule
        autoregressive    & \xmark                  & \xmark                  & \cmark                         & \cmark                          \\
        \bottomrule
    \end{tabular}
    \midsepdefault

    \caption{Comparison between all four algorithms: QAOA, PG-QAOA, CD-QAOA and \agentname.}
    \label{table:comparison}
\end{table}

\section{\label{sec:algo}Mixed Discrete-Continuous Policy Gradient using Deep Autoregressive Networks}

Although RL is used as an optimizer in both PG-QAOA and CD-QAOA, it serves two fundamentally different purposes. In PG-QAOA it is employed for continuous optimization of the protocol durations $\{\alpha_j\}$, while in CD-QAOA it is used to find the solution to the discrete combinatorial task of ordering the unitaries in the protocol sequence. In this section, we illustrate how to combine the two aspects together into a unified RL-based algorithm.

We have seen that with the help of RL one can tremendously  enhance the properties of the QAOA ansatz in very different ways, cf.~Table~\ref{table:comparison}:
For instance, PG-QAOA has the important desired property that it is robust to noise. Moreover, it does a completely gradient-free optimization of the continuous protocol durations. On the other hand, CD-QAOA, enhances the variational ansatz itself by offering the appealing ability to select the order in which three or more unitaries can be applied in the protocol sequence. Moreover, it also introduces an autoregressive deep neural network to encode causality (i.e., which unitary is optimal at a given episode step depends on the unitaries chosen hitherto).
The imminent question arises as to whether we can design an algorithm which makes the best of both worlds.

\subsection{\label{subsec:hybrid}Autoregressive Policy Ansatz for Hybrid Discrete-Continuous Action Spaces}

Recently, a number of studies have considered the problem of simultaneous discrete/continuous control using RL~\cite{kulkarni2016hierarchical, fan2019hybrid, wei2018hierarchical, hausknecht2015deep, delalleau2019discrete, bester2019multi, xiong2018parametrized, fan2019hybrid, wei2018hierarchical, neunert2020continuous}.
Following the notation of Ref.~\cite{masson2015reinforcement}, we describe the RL problem within the framework of parametrized-action Markov decision processes (PAMDPs). The major difference, compared to ordinary MDPs, is the definition of the action space:
$
    \mathcal A = \bigcup_{\ad\in \Ad, \ac \in \Ac} (\ad, \ac), \;
    \Ad = \{H_j\}_{j=1}^{\vert \Ad \vert},\;
    \Ac=[0,1],
$
where $\vert \Ad \vert$ denotes the cardinality of the discrete action set. As before, the state space contains all possible sequences of actions, and the reward is the (negative) energy density of the quantum state, given once at the end of the protocol.

In this section, we present a unified continuous-discrete quantum control algorithm, called \agentname, based on a hybrid policy which optimizes simultaneously the discrete and continuous degrees of freedom in the policy. The policy can be decomposed as a product of two coupled auxiliary policies -- one for the continuous actions, $\pc$, and the other for the discrete actions, $\pd$:
\begin{equation}
    \pi_{\params}(\tau) =\pc\left(\tau^{\mathrm c} \right) \pd\left(\tau^{\mathrm d}\right),
\end{equation}
where $\tau^{\nu} = (a^{\nu}_1,\dots,a^{\nu}_q),\; \nu\in\{\mathrm c,\mathrm d\}$ defines the discrete/continuous subsequence of actions in each trajectory of length $q$.
Denoting, as before, the RL state by $s_j\eq (a_{1}, \cdots, a_{j-1})$ with the hybrid action $a_\ast \eq (\ac_\ast, \ad_\ast)$, we define a generalized continuous/discrete autoregressive model for the policy, following Eq.~\eqref{eq:autoreg_pi}. Adopting the short-hand notation
$\pi^\nu_{\boldsymbol{\theta}}\left(a_{j}^\nu \mid s_j\right)\eq \pi^\nu_{\boldsymbol{\theta}}\left(a_{j}^\nu\mid a_{1}, \cdots, a_{j-1} \right)$, the policy can be written as
\begin{equation}
    \pi_{\boldsymbol{\theta}}\left(a_{1}, a_{2}, \cdots, a_{q}\right)\eq \prod_{j\eq 1}^{q} \pd\left(\ad_{j}\mid s_j\right)\pc\left(\ac_{j} \mid s_j, \ad_{j}\right).
\end{equation}
As expected, at every step $j$, the action $\ac_j$ is sampled from a continuous distribution, whose parameters depend on the discrete action $\ad_j$ selected at the same step $j$. This is natural, since different discrete actions may require different corresponding continuous distribution parameters $\kappa,\xi$.

Additionally, similar to CD-QAOA, we impose a further restriction that no discrete action can occur in the trajectory consecutively. We use a Sigmoid-Gaussian distribution to bound the samples for the continuous actions, and normalize the durations $\alpha_j\propto \ac_j\sim\pc$ to fix the total protocol duration to $\sum_{j=1}^q \alpha_j = T$; using the Beta distribution instead results in a similar performance.

\subsection{\label{subsec:hybrid}Deep Autoregressive Policy Network}

\begin{figure}[t!]
    \centerline{
        \includegraphics[width=0.7\textwidth]{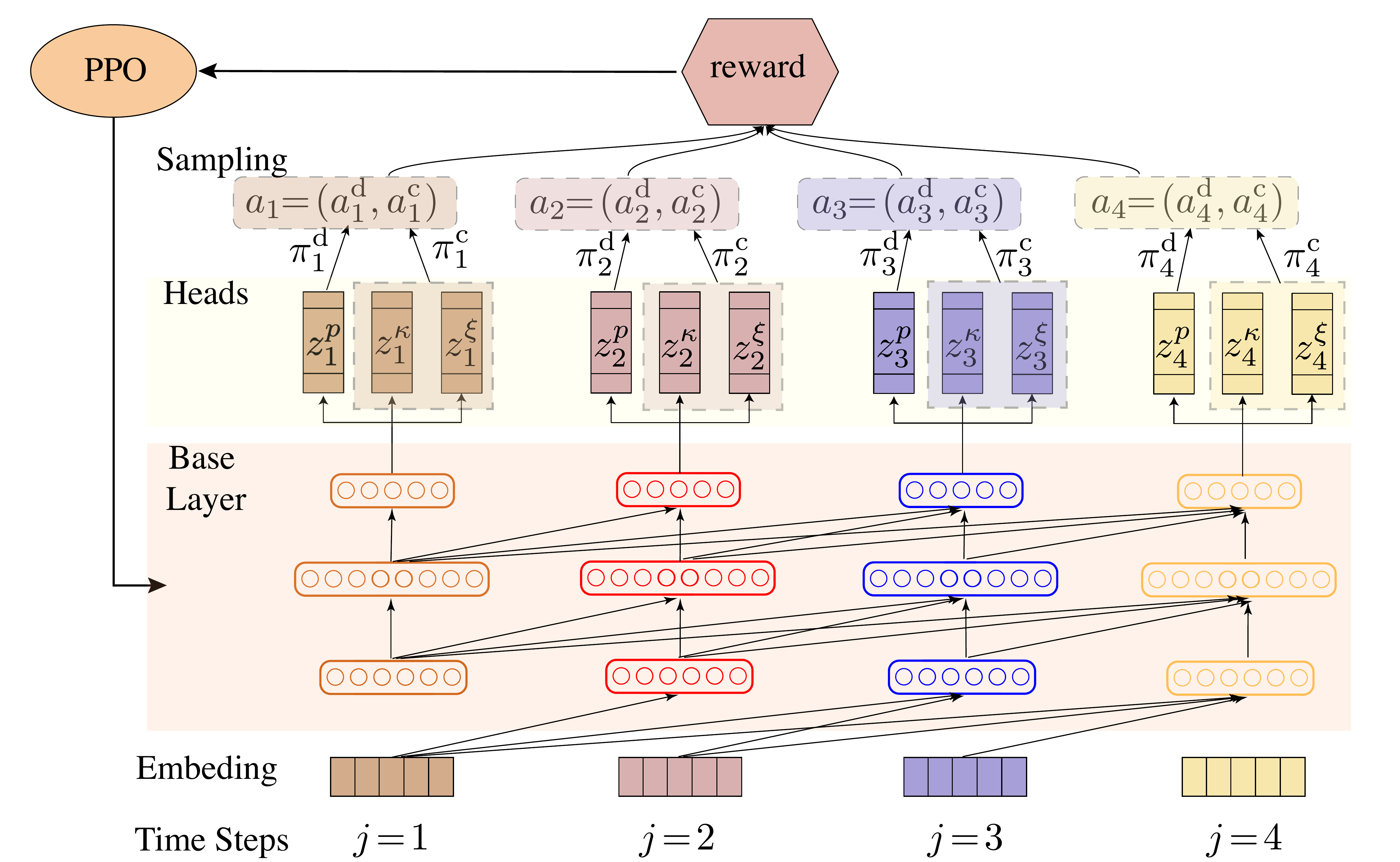}}
    \caption{
        \label{fig:arnet}
        \small
        Schematic representation of \agentname{} and the deep autoregressive network for $q=4$ (see text).
        The time step $j$ also corresponds to the gate index.
        The policy network is composed of
        (i) an embedding layer to encode the continuous and discrete actions as input.
        (ii) The base layer implements the causal autoregressive structure (see arrows).
        (iii) The heads are three-fold, one for the discrete distribution parameters, and two for the continuous distribution parameters.
        A batch of actions are sampled to evolve the quantum state and compute the negative energy density as a reward. Proximal hybrid Policy Optimization (PPO) is used to update the policy network.
        The pseudocode for \agentname{} is shown in Algorithm~\ref{alg:ARPG}.
    }
    \vspace{-1em}
\end{figure}

We implement the policy ansatz variationally, using a deep neural network called the policy network. In Fig.~\ref{fig:arnet}, we show a cartoon of the model for illustration purposes.
The network consists of base layers with intermediate output $\boldsymbol{y}$, followed by three independent head layers with outputs $\boldsymbol{z}^p, \boldsymbol{z}^\kappa, \boldsymbol{z}^\xi$, respectively. The three heads learn the discrete probability distribution $\pi^\mathrm d$, and the parameters $\kappa,\xi\in\mathbb{R}^+$ which define the continuous probability distribution $\pi^\mathrm c$. Each head outputs a vector of size $|\Ad|$ -- so that the model can learn a set $(\kappa,\xi)$ for every distinct discrete action. Notice that each head output depends on the joint base layer parameters $(\boldsymbol{W},\boldsymbol{b})$, but not on the parameters $(\boldsymbol{V},\boldsymbol{c})$ of any of the other two heads; thus, the base layers are shared by all three heads.
In practice, we find that a base layer, comprised of two hidden layers, can already achieve a good performance; one can in principle add more layers for enhanced expressivity.

The above description focuses on a single episode step $j$ out of a total of $q$ steps in an episode.
The autoregressive feature of the ansatz can then be built-in, by allowing the outputs of the base layers from previous steps to become inputs into the layers at subsequent episode steps [Fig.~\ref{fig:arnet}].

Let us denote the input to the autogressive network by $(x_1, x_2, \cdots, x_q)$, and the weights and bias parameters of the base layer by $W_j\in \mathbb R^{ d_{\text{h}} \times (i-1)|\Ad|}$ and $b_j \in \mathbb R^{ d_{\text{h}} }$, respectively, where $d_{\text{h}}$ is the hidden dimension. Then, the intermediate output $(y_1, y_2, \cdots, y_q)$ of the base layer reads as
\begin{equation}
    y_j\eq g(W_{j}x_{<j} + b_{j}),\quad j\eq 1, 2, \cdots q,
\end{equation}
where $x_{<j}=\left(x_{j-1}, \cdots, x_{1}\right)^{T}\in\mathbb R^{ (j-1)|\Ad|}$ denotes the input of all previous steps preceding step $j$; for $j\eq 1$, we set $W_jx_{<j} + b_j\eq b_j$.\footnote{In practice, implementing the autoregressive constraint $_{<j}$ can be achieved using masks (one for each set of weights).}
We use ReLU nonlinearities $g(\cdot)$.

The output of the base layer $(y_1, y_2, \cdots, y_q)$ can be viewed as an input to the three-head layer. The three-head layer contains three heads with independent
weights $V_{j}^p, V_{j}^\kappa, V_{j}^\xi \in \mathbb R^{ |\Ad| \times j d_{\text{h}}}$
and biases $c^p_j, c^\kappa_j, c^\xi_j \in \mathbb R^{ |\Ad| }$.
The three-head layer output, $(z_1, z_2, \cdots, z_q)$, are the parameters for the discrete and continuous distributions: $z_j^p$ are the categorical distribution parameters; $z_j^\kappa$ and $z_j^\xi$ are the two parameters for the sigmoid-Gaussian distribution [cf.~App.~\ref{app:sigmoid-gaussian-1}]:
\begin{equation}
    z_j^p \eq \log\left(\operatorname{SoftMax}( V_{j}^p y_{\leq j} + c^p_j)\right), \qquad
    z_j^\kappa \eq V_{j}^\kappa y_{\leq j} + c^\kappa_j, \quad
    z_j^\xi \eq \exp( V_{j}^\xi y_{\leq j} + c^\xi_j),
\end{equation}
where $y_{\leq j}=\left(y_{j}, \cdots, y_{1}\right)^{T}\in\mathbb R^{ j d_{\text{h}}}$.\footnote{Note that here we are able to use the "$=$" sign because the previous layer of operation has already filtered out the "$=$" sign for those steps.} To define a categorical distribution, we use a SoftMax\footnote{Note that this function is not operated element-wise like the others; it is applied on the whole vector of dimension $ |\Ad|$).} nonlinarity: $\operatorname{SoftMax}(v)[i] \eq \exp(v[i]) / \sum_{k=1}^{|\Ad|}\exp(v[k])$, where $v\eq V_{j}^p y_{\leq j}+c^p_j \in \R^{|\Ad|}$, and $[\cdot]$ takes the index;
we learn the log-probability to achieve a resolution over a few orders of magnitude, and to stabilize the learning process.

We apply ancestral sampling to draw actions from the autoregressive policy. Starting from the heads layer at step $j=1$, we first sample $\ad_1 \sim \pi(\ad_1) \eq \operatorname{Categorical}(\exp(z_1^p))$; we use the sampled discrete action $\ad_1$ to look-up the corresponding parameters $\kappa =z_1^\kappa[\ad_1]$ and $\xi=z_1^\xi[\ad_1]$\footnote{Here, $[\ad_1]$ means taking the component by index.} for the continuous action distribution.
Then we sample the duration $\alpha_1  \propto \ac_1 \sim \pi(\ac_1 | \ad_1) \eq \mathcal{SN}\left(z_1^\kappa[\ad_1], {(z_1^\xi}[\ad_1])^2\right)$. The sampling output is passed as an input at the second step $j=2$. To do this, we use as an embedding for $(\ad_1, \ac_1)$ represented by the variable $x_1$, where
$x_1[i] = \ac_1  \text { if }  i\eq \ad_1$, and $x_1[i] =0$ otherwise.
Going on, we repeat the process: we sample successive actions $\ad_2, \ac_2\sim \pi(\ad_2 |x_1), \pi(\ac_2 |x_1, \ad_2)$. The sampling, or forward pass, through the network is then repeated $q$ times, until we reach the end of the episode; thus, at step $j$ we have $\ad_j, \ac_j\sim \pi(\ad_j |x_{<j}), \pi(\ac_j |x_{<j}, \ad_j)$. This gives the trajectory $\tau$ of mixed discrete-continuous actions. Note that the time complexity of the process is $\mathcal{O}(q \times |\Ad|)$.

\subsection{\label{subsec:hybrid}Proximal Hybrid Policy Optimization}

The set of all weights and biases, $\params\eq\{W_{j},  b_j,  V_{j}^p, V_{j}^\kappa, V_{j}^\xi, c^p_j, c^\kappa_j, c^\xi_j\}_{j\eq1}^{q}$, defines the learnable parameters of the autoregressive policy network. We now discuss how to compute the policy gradients and define an update rule for $\params$.

Our goal is to maximize the RL objective within the trust region~\cite{schulman2015trust} for the continuous and discrete policy
\begin{equation}
    \label{eq:PPO}
    \mathbb{E}_{\tau}\left[\frac{\pi_{\params}\left(\tau\right)}{\pi_{\params_{\text {old}}}\left(\tau\right)} {A_{\params_\mathrm{old}}(\tau)}\right],
    \quad\text{subject to}\quad
    {\mathbb{E}}_{\tau}\left[\operatorname{D_{KL}}\left[\pi^\nu_{\params_{\text {old }}}\left(\cdot \right), \pi^\nu_{\params}\left(\cdot \right)\right]\right] \leq \delta^\nu,
\end{equation}

\noindent
where $\E_\tau[\ \cdot\ ]$ is a shorthand notation for $\E_{\tau\eq(a_1, \cdots, a_q)\sim \pi_{\bf \params_\mathrm{old}}}[\ \cdot\ ]$. The Kullback–Leibler (KL) divergence is defined as
$\operatorname{D_{\text{KL}}}(\pc, \pto{c})\eq \int_{x \in \Ac} \pc(x) \log \left(\frac{\pc(x)}{\pto{c}(x)}\right) \dd{x}$,
and similarly for $\nu\eq \mathrm d$;
$\delta^\nu$ defines a constraint on the size of the discrete policy or continuous policy updates in distribution space. Here, $\params_{\textrm{old}}$ denotes the parameters before the update; $A_{\params_\mathrm{old}}(\tau)= R(\tau) - b$ is the advantage function -- the return (negative energy density) for a given  trajectory w.r.t.~the baseline $b$.

In practice, we utilize a clipped surrogate RL objective~\cite{schulman2017proximal} with two clipping parameters $\epsilon^\nu$.
The idea is to update the continuous and discrete policies using different $\epsilon^\nu$ during policy optimization. This allows for the discrete policy $\pd$ to change more quickly/more slowly as compared to the continuous policy $\pc$. Hence, the hybrid PPO RL objective reads as
\begin{equation}
    \mathcal J(\params)=
    \mathbb{E}_{\tau} \bigg[  \mathcal G^\mathrm d(\td; \params, \epsilon^\mathrm d) +  \mathcal G^\mathrm c(\tc; \params, \epsilon^\mathrm c)\bigg] + \beta^{-1}_{{S}} (\SD + \SC) ,
    \label{eqn:ppo-maxent}
\end{equation}
with
\begin{equation}
    \mathcal G^\nu(\tau^\nu; \params, \epsilon^\nu) = \min \bigg\{
    \rho_\params^\nu (\tau^\nu) A^\nu_{\params_\mathrm{old}}(\tau^\nu),\;
    \operatorname{clip}\left(\rho_\params^\nu(\tau^\nu), 1-\epsilon^\nu, 1+\epsilon^\nu\right) A^\nu_{\params_\mathrm{old}}(\tau^\nu)  \bigg\},
    \label{eqn:ppo-adv}
\end{equation}
where $\rho_\params^\nu (\tau^\nu)\coloneqq\frac{\pi^\nu_{\params}(\tau^\nu)}{\pi^\nu_{\params_{\mathrm{old}}}(\tau^\nu)}$ is the importance weight ratio of two policies associated with trajectory $\tau^\nu$. The clip function, defined as $\mathrm{clip}(\rho, x,y)= \max \big( \min \left(\rho, x \right) , y\big)$ sets the value of $\rho_\params$ to be within the interval $[x, y]$, and constrains the likelihood ratio from Eq.~\eqref{eq:PPO} to the range $[1-\epsilon, 1+\epsilon]$.
The entropy terms [right-most part of Eq.~\eqref{eqn:ppo-maxent}] are discussed below.
Our goal is to find those parameters $\params$ which maximize $\mathcal J(\params)$.

To understand the hybrid PPO algorithm, consider two limiting cases first. In the extreme case when $\epd\to0$,~i.e.~the discrete policy $\pd$ is kept fixed, our algorithm reduces to PG-QAOA.
On the other hand, when $\epc\to0$, the continuous policy is kept fixed; if this fixed policy additionally corresponds to the greedy ``expert policy'' defined by the Powell optimizer, the algorithm is reduced to CD-QAOA.
In this sense, for finite values of $\epc,\epd>0$, \agentname{} can be viewed as a smooth interpolation between PG-QAOA and CD-QAOA.

In order to incentivize the agent to explore the action space during the early stages of training, we also added entropy to the RL objective, cf.~Eq.~\eqref{eqn:ppo-maxent}. The entropy for a discrete/continuous policy is defined as   $\SD(\pi^\mathrm d)\eq -\sum_{x \in \mathcal{X}} \pi^\mathrm d(x) \log \pi^\mathrm d(x)$ or $\SC(\pi^\mathrm c)\eq -\int_{x \in \mathcal{X}} \pi^\mathrm c(x) \log \left( \pi^\mathrm c(x) \right) \dd x$, respectively.
The coefficient $\beta^{-1}_S$ in Eq.~\eqref{eqn:ppo-maxent} defines an effective temperature, which we anneal with increasing the number of iterations. It is easy to see that the total entropy $\mathcal S= \SD +\SC$ associated with the hybrid policy consists of
a discrete
$
    \SD  = \sum_{j=1}^q \E_{a_{<j}\sim \pi_\params} \mathcal S^\mathrm d\big(\pd(\;\cdot\; | a_{<j})\big)$,
and a continuous
$
    \SC  = \sum_{j=1}^q\E_{a_{<j}\sim \pi_\params, \ad_j\sim \pd} S^\mathrm c\big(\pc(\;\cdot\; |a_{<j}, \ad_j)\big)
$
contribution.
The RL agent has to maximize the total expected return while also maximizing the entropy associated with the policy.

In RL, there are two common ways to incorporate entropy in practice~\cite{levine2018reinforcement}:
(i) whenever one can compute a closed-form expression for the entropy, entropy is added as a separate term to the objective which can be thought of as entropy regularization. Note that it is the autoregresssive structure that makes it possible to obtain the exact value for the entropy $\SD$: for $\pd(\;\cdot\; | a_{<j}) \eq \operatorname{Categorical}(\exp (z_j^p))$, the entropy is $\SD\big(\pd(\;\cdot\; | a_{<j})\big)\eq - \sum_{k\eq 1}^{|\Ad|}z_j^p[k] \cdot\exp(z_j^p[k])$.
(ii) Often times it is not possible to compute the value for the entropy, since the expression is not analytically tractable; in such cases, the maximum entropy formulation~\cite{haarnoja2018soft, haarnoja2018soft2, haarnoja2017reinforcement} still allows us to add to the reward a sample estimate of the entropy, known as an entropy bonus: $ R^\mathrm c(\tau)\leftarrow  R^\mathrm c(\tau)+\beta^{-1}_S \E_{\ac \sim \pc} \big[- \log \pc \big ]$. In this study, we add an entropy bonus to take into account the entropy of the continuous policy $\pi^c$.

\section{Application: Quantum Ising Model in the Presence of Noise} \label{sec:Ising-1/2}

To test the performance of \agentname, we investigate the ground state preparation problem for a system of $N$ interacting qubits (i.e.~spin-$1/2$ degrees of freedom), described by the Ising Hamiltonian introduced in Eq.~\eqref{eq:IM}. We use periodic boundary conditions and work in the zero momentum sector of positive parity, which contains the antiferromagnetic ground state. We emphasize that this model is non-integrable, i.e., it does not have an extensive number of local integrals of motion; as a consequence, no closed-form analytical description is known for its eigenstates and eigenenergies. Moreover, the lack of integrability results in chaotic quantum dynamics.

In the following, $J\!=\!1$ sets the energy unit, $h_z/J\!=\!0.4523$ and $h_x/J\!=\!0.4045$. In the thermodynamic limit, $N\to\infty$,
these parameters are close to the critical line of the model, where a quantum phase transition occurs in the ground state between an antiferromagnet and a paramagnet; for the finite system sizes we can simulate, the critical behavior is smeared out over a small finite region. In Ref.~\cite{matos2020quantifying}, using QAOA it was shown that this region of parameter space appears most challenging in the noise-free system.

We initialize the system in the $z$-polarized product state $|\psi_i\rangle\!=\!|\!\!\uparrow\cdots\uparrow\rangle$, and aim to prepare the ground state of $H$. We use the negative energy density $-\mathcal{E}=-E/N$ as a reward for the RL agent, cf.~Eq.~\eqref{eqn:qaoa}, which is an intensive quantity as the number of qubits $N$ increases. In this study, we are mostly interested in exploring the behavior of the system, subject to various kinds of noise/uncertainty.
Our primary focus is quantifying the effects of noise on the achievable fidelity, w.r.t.~the noise-free values.
We deliberately select a fixed duration of $JT\eq10$ far from the adiabatic regime, such as to exhibit the benefits of the CD-QAOA ansatz over QAOA [cf.~App.~\ref{app:durations}].

We point out that, working at a fixed duration $T$, it is not always possible to achieve high-fidelity ground states. This is easy to see for decoupled qubits, where the magnitude of the spin precession frequency on the Bloch sphere (so-called Larmor precession frequency) is set by the fixed strength of the magnetic field $(h_x,0,h_z)$: hence, fixing the total protocol duration $T$, it may be physically impossible to reach the target state in the allotted time. This behavior leads to the notion of the quantum speed limit (QSL) -- the minimum time required to prepare the ground state with unit fidelity.

\subsection{Three Noise Models} \label{subsec:noise}

When operating present-day quantum devices, one is confronted with various sources of uncertainty. Since the exact form and details depend on the peculiarities and particularities of the underlying experimental platform, it is desirable to construct algorithms, capable of learning such details without extra human input.
In this study, our RL agent learns in a simulator. To mimic the diversity of uncertain processes that can occur, we consider three types of noise.

\subsubsection{Classical Measurement Gaussian Noise}

Noise naturally occurs due to imperfect measurements. For instance, the measurement signal is often present in a form of currents and voltages, whose values can only be determined within the resolution of the measurement apparatus. In practice, experimentalists perform a large number of measurements and average the result in the end to obtain an estimate for the value of an observable. By the central limit theorem, in the limit of large sample sizes, the statistics of the measurement data is approximated by a Gaussian distribution. To model this behavior, we use small Gaussian noise to add uncertainty in the reward signal:
$
    \mathcal{E}_\gamma (\{\alpha_i\}_{i=1}^q, \td)= \mathcal{E}(\{\alpha_i\}_{i=1}^q, \td) + \epsilon_\gamma
$,
where $\epsilon_\gamma \sim \mathcal N (0, \gamma^2)$. %

\subsubsection{Quantum Measurement Noise}

In quantum mechanics, there is another, intrinsic, kind of noise, which arises due to the quantum nature of the controlled system. Consider the evolved state $|\psi(T)\rangle = U(\{\alpha_j\}_{j=1}^q ,\tau)|\psi_i\rangle$ at the end of the protocol. The expected measurement for the energy density $\mathcal{E}=N^{-1}\langle \psi(T)|H|\psi(T)\rangle$ is obtained within a \textit{quantum} uncertainty,
$
    \Delta \mathcal{E} = N^{-1} \sqrt{ \langle \psi(T)|H^2|\psi(T)\rangle - \langle \psi(T)|H|\psi(T)\rangle^2 }
$,
set by the energy variance in the final state. In the limit of a large number of measurements, quantum noise can be simulated using a Gaussian distribution
$
    \mathcal{E}_Q (\{\alpha_i\}_{i=1}^q, \td)= \mathcal{E}(\{\alpha_i\}_{i=1}^q, \td) + \epsilon_Q
$,
where $\epsilon_Q \sim \mathcal N (0, \Delta \mathcal{E}^2)$. Note that the width of the Gaussian depends on the final state $|\psi(T)\rangle$: in the early stages of training, $|\psi(T)\rangle$ is typically far away from any of the eigenstates of $H$; therefore, the energy variance $\Delta \mathcal{E}$ will be large and finite. However, towards the later training stages, when the agent learns to prepare a state close to the target ground state, the energy variance will go down. Hence, one can think of the quantum noise as a Gaussian noise with a time-dependent strength.

\subsubsection{Noise arising from Gate Rotation Errors}

Finally, we also consider the uncertainty in implementing the unitaries $U_i$. We focus on gate rotation errors~\cite{sung2020exploration}, caused by imperfections in the durations $\alpha_i$:
$
    \mathcal{E}_\delta (\{\alpha_i\}_{i=1}^q, \tau)= \mathcal{E}(\{\alpha_i + \epsilon_{i}\}_{i=1}^q)
$,
where $\epsilon_{i} \sim \mathcal N (0, \delta^2)$.
This defines a simplified error model for coherent control, an important source of errors in present-day state-of-the-art quantum computing hardware~\cite{arute2019quantum}, and which is especially pertinent to the case of quantum computers which are utilized frequently but calibrated only periodically.

\section{Numerical Experiments and Results} \label{sec:experiment}

\begin{figure}[t!]
    \centering
    \hspace{2em}
    \begin{tikzpicture}
        \begin{axis}[%
                line width=2.5pt,
                hide axis,
                xmin=10,
                xmax=50,
                ymin=0,
                ymax=0.4,
                legend style={draw=none, legend columns=-1}
            ]
            \addlegendimage{sblue}
            \addlegendentry{QAOA \quad \quad}
            \addlegendimage{sorange}
            \addlegendentry{PG-QAOA \quad \quad};
            \addlegendimage{sgreen}
            \addlegendentry{CD-QAOA \quad \quad};
            \addlegendimage{sred}
            \addlegendentry{\agentname \quad \quad};
        \end{axis}
    \end{tikzpicture}
    \\
    \vspace{-0.5em}
    \includegraphics[width=1.0\textwidth]{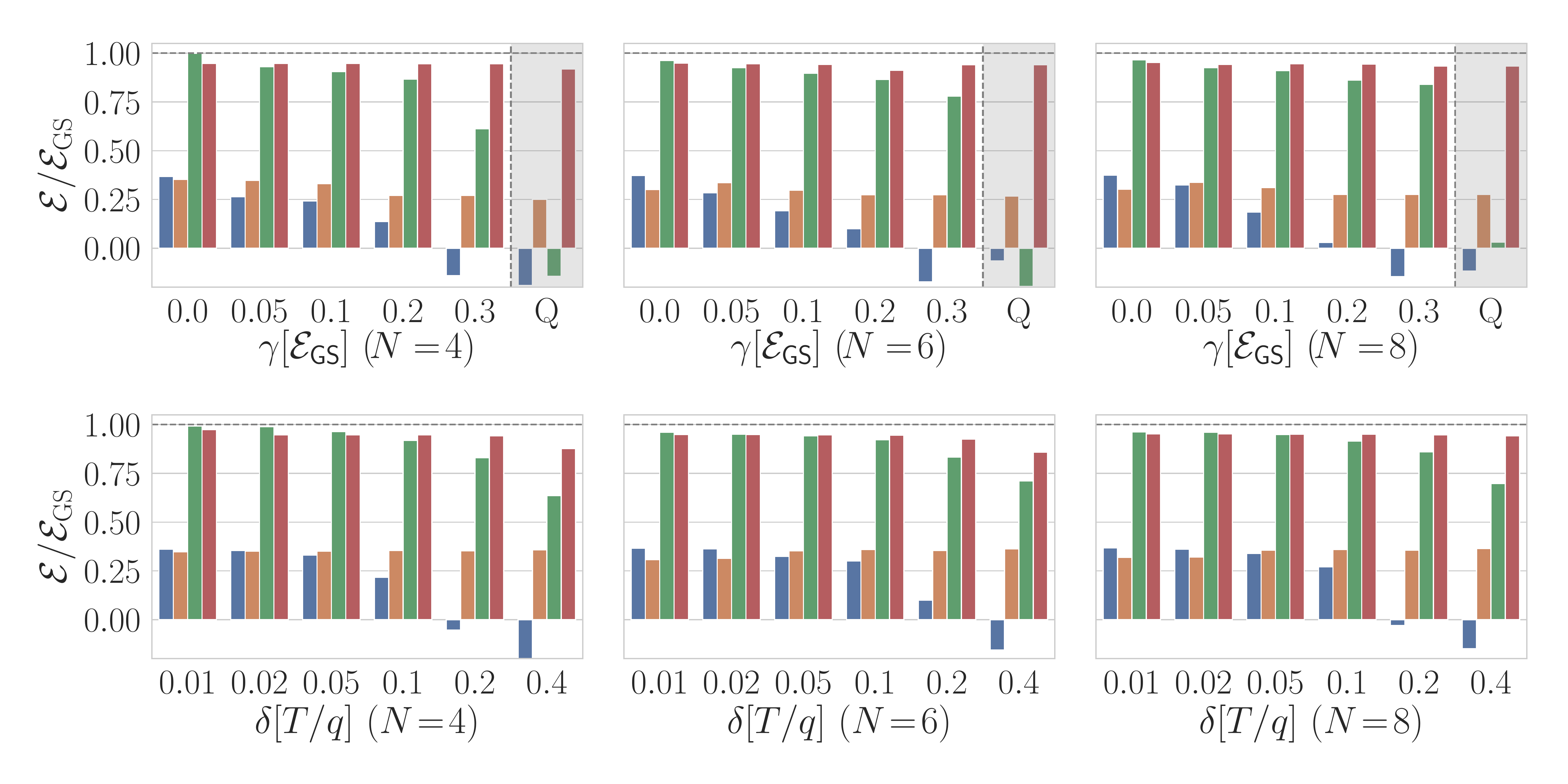}
    \caption{
        \small Energy minimization against different noise levels with circuit depths $p\eq q/2\eq4$ and protocol duration $JT\eq 10$ for four different optimization methods: QAOA, PG-QAOA, CD-QAOA (red line), \agentname.
        The initial and target states are $|\psi_i\rangle\!=\!|\!\!\uparrow\cdots\uparrow\rangle$ and $|\psi_\ast\rangle\!=\!|\psi_\mathrm{GS}(H)\rangle$ for $h_z/J=0.4523$ and $h_x/J=0.4045$.
        The alternating unitaries for conventional QAOA and PG-QAOA are generated by $\Ad=\{ H_1, H_2\}$;
        for CD-QAOA and \agentname, we extend this set using adiabatic gauge potential terms to $\Ad=\{ H_1, H_2; Y, X|Y , Y|Z \}$. The system sizes are $N\!=\!4, 6, 8$.
    }
    \label{fig:noiseT10}
\end{figure}

To evaluate the performance of the trained agent, we eliminate the uncertainty associated with the probabilistic nature of the policy: we take the discrete action which maximizes the categorical distribution $\pi^\mathrm d$, and only keep the mean of the continuous distribution $\pi^\mathrm c$, setting its width to zero. This defines a natural greedy policy to test the ability of the RL agent.

We performed a number of numerical experiments to study the effect of the noise on the performance of the four algorithms QAOA, PG-QAOA, CD-QAOA and \agentname, for the three different sources of uncertainty: classical and quantum measurement noise, and gate rotation noise. We vary both the noise strength, and we look at three different system sizes for two protocol durations each. The results of these experiments can be summarized, as follows.

Figure~\ref{fig:noiseT10} shows the best achievable energy at a protocol duration $JT=10$ against different noise types and system sizes: the top row shows data for various measurement noise strengths, with the shaded area marking the special case of quantum noise; the noise strength is measured in percentages of the achievable ground state energy density: e.g., a noise strength of $\gamma=0.3$ corresponds to an average deviation from the actual energy of about $30\%$.
The bottom row displays results when varying the gate noise strength. Here, the noise strength is defined as a percentage of the mean gate duration $T/q$. The three columns correspond to system sizes $N=4$ (left), $N=6$ (middle) and $N=8$ (right).

When $T<T_\mathrm{QSL}$ is chosen below the QSL, we find that QAOA and PG-QAOA fail to reach the ground state in the time allotted, as a result of having an overconstrained control space $\Ad=\{H_0,H_1\}$. Nonetheless, the noise-robust character of PG-QAOA becomes pronounced at increased values of the noise strength. Since the initial quantum state is far away from the target ground state, the best ratio $E/E_\mathrm{GS}$ found by QAOA can even be negative. The $JT=10$ duration exhibits the advantage of using the generalized QAOA ansatz brought in by CD-QAOA: suitably enlarging the discrete action space $\Ad = \{H_1,H_2,Y,X|Y, Y|Z\}$ unlocks paths in Hilbert space which are inaccessible to QAOA. Hence, CD-QAOA and \agentname{} find the largest rewards in the noise-free case. A large noise strength reduces visibly the ability of CD-QAOA to find the ground state, with the performance being particularly bad for quantum measurement noise (Q). However, the hybrid policy optimizer allows \agentname{} to emerge as a noise-robust algorithm, agnostic to the source of noise applied to the system.

\section{\label{sec:outro}Conclusion and Outlook}

In summary, we presented \agentname{} -- a versatile and noise-robust quantum control algorithm based on the QAOA variational ansatz. The algorithm inherits valuable features from its ancestors:
(i) the noise-robust property of PG-QAOA allows us to find optimal durations probabilistically.
(ii) the generalized QAOA ansatz of CD-QAOA makes it possible to select the order in which a set of unitaries appears in the control sequence. While we focused on physically-motivated unitaries, we emphasize that the ansatz is completely general and applicable to a large variety of unitaries/quantum gate sets useful for both theoretical and experimental studies.
We had to modify these ``ancestors'' accordingly: in PG-QAOA we  introduced a mechanism to fix the total protocol duration, and introduced a stochastic policy based on the compactly supported Beta function; in CD-QAOA we changed the low-level optimizer to gradient-free Powell, as opposed to the gradient-based SLSQP which did not give a reasonable performance in the presence of noise.
\agentname{} extends PG-QAOA and CD-QAOA with both the use of a generalized autoregressive architecture which incorporates the parameters of the continuous policy, and the derivation of an extension of Proximal Policy Optimization applicable to hybrid continuous-discrete policies.

We tested the performance of \agentname{} using the unitary dynamics of a quantum Ising chain subject to various sources of noise: classical and quantum measurement noise as well as uncertainty leading to errors in the application of quantum unitary gates. In particular, we demonstrated that \agentname{} successfully outperforms its ancestors in the highly-constrained non-adiabatic regime, irrespective of the noise model selected.
Thus, \agentname{} is not only noise-robust but also agnostic to the physical source of noise. This opens up the exciting possibility of using machine learning to `learn' the particularities of noisy experimental environments, which often depend on the chip architecture and can even change in the course of exploitation.
However, the presented results are obtained using numerical simulations based on certain theoretical assumptions; it remains to test the performance of \agentname{} on realistic noisy intermediate-scale quantum computing devices.

The \agentname{} is a versatile method that can be extended along several directions. For instance, the current version of \agentname{} defines a fixed sequence/protocol length. However, the algorithm is versatile enough to accommodate a variable length of the protocols after a slight modification. To do so, one can simply add a ``stop'' action to the discrete action set $\mathcal{A}^d$. If the agent happens to choose the stop action, then the episode comes to an end immediately and we measure the energy of the evolved quantum state.

There also exist a number of exciting alternatives for the policy network architecture to explore. Although it has to incorporate temporal causality, notice that the architecture is not limited to the autoregressive choice used in this study; e.g., it can be generalized to a recurrent neural network (RNN), a Long Short Term Memory (LSTM) network, or a transformer with the attention mechanism~\cite{vaswani2017attention} and all its modern variants~\cite{kitaev2020reformer, choromanski2020rethinking, wang2020linformer, tay2020efficient}. In the present study, we chose the autoregressive network for its sheer simplicity.
Moreover, the continuous policy head can be generalized to capture distributions with more than two modes using the normalizing flow method, which would additionally boost the expressivity of the policy~\cite{tang2018boosting}.

\acks{We wish to thank Vitchyr Pong for valuable discussions. This work was partially supported by the Department of Energy under Grant No. DE-AC02-05CH11231 and No. DE-SC0017867 (L.L., J.Y.), and by the National Science Foundation under the NSF QLCI program through grant number OMA-2016245 (L.L.). M.B.~was supported by the Bulgarian National Science Fund within National Science Program VIHREN, contract number KP-06-DV-5, and the Marie Sklodowska-Curie grant agreement No 890711.

    We used Powell methods~\cite{powell1964efficient} implemented in SciPy~\cite{2020SciPy-NMeth} for the QAOA solver; we used  W\&B~\cite{wandb} to organize and analyze the experiments. The reinforcement learning networks are implemented in NumPy\cite{harris2020array}, and   TensorFlow~\cite{tensorflow2015-whitepaper} and TensorFlow Probability~\cite{dillon2017tensorflow}; the quantum systems are simulated in \href{https://github.com/weinbe58/QuSpin\#quspin}{Quspin}~\cite{weinberg2017quspin, weinberg2019quspin}. We thank Berkeley Research Computing (BRC) for providing the computational resources.}

\clearpage
\bibliography{msml2021}

\appendix
\newpage

\section{Pseudocode and Algorithm Hyperparameters}

The pseudocode for \agentname{} is outlined in Algorithm~\eqref{alg:ARPG}. The agent samples a batch of actions from the autoregressive network. Then, the corresponding expected energy density is computed using a classical simulator for the quantum dynamics.
Below, we focus on the noise-free case; dealing with noise requires a trivial modification following Sec.~\ref{subsec:noise}.
The baseline for the reward is estimated through an exponential moving average. Finally, proximal policy optimization is applied to update the agent's policy.

\begin{algorithm}[!thbp]
    \caption{Autoregressive network based reinforcement learning: \agentname}
    \label{alg:ARPG}
    \begin{algorithmic}[1]
        \REQUIRE batch size $M$, learning rate $\eta_t$, total number of iterations $T_{\mathrm{iter}}$, exponential moving average coefficient $m$, entropy coefficient $\beta^{-1}_{{S}}$, PPO gradient steps $K$.\\
        \STATE Initialize the autoregressive network and initialize the moving average $\hat R\eq 0$.
        \FOR {$t=1,..,T_{\mathrm{iter}}$}
        \STATE Autoregrssively sample batch $B$ hybrid actions of size $M$:
        \vspace{-0.6em}
        $$a_{1}^{\{k\}}, a_{2}^{\{k\}}, \cdots, a_{q}^{\{k\}} \sim \pi_{\boldsymbol{\theta}}\left(a_{1}, a_{2}, \cdots, a_{q}\right),  \ k = 1, 2, \cdots, M.$$
        \vspace{-1.5em}
        \STATE Measure the observables and use the negative energy density as the return  and compute the moving average of the return
        \vspace{-1.5em}
        $$R_k = -\mathcal E_k= - \frac{1}{N} \langle\psi_i|U^\dagger(\{a_j^{\{k\}}\}_{j=1}^q ) H U(\{a_j^{\{k\}}\}_{j=1}^q ) |\psi_i\rangle, \quad \hat R = m \cdot \hat R + (1-m) \cdot \frac{1}{M}\sum_{k=1}^M R_k.$$
        \vspace{-1.5em}
        \STATE Compute the advantage estimates
        $ A_k = R_k - \hat R$
        \FOR {$k=1,..,K$}

        \STATE Evaluate the samples' likelihood using the parameter from the last iterations, i.e. $\pto{}(a_{1}^{\{k\}}, a_{2}^{\{k\}}, \cdots, a_{q}^{\{k\}})$ and compute the importance weight $\rho_k^\nu$
        \STATE Using the advantage estimation and importance weight to compute $\mathcal G^d_k , \mathcal G^c_k, \SD_k , \SC_k$.
        \STATE Compute the \agentname{} objective Eq~(\ref{eqn:ppo-maxent}) and backpropagate to get the gradients.
        \[ \nabla_{\params}\mathcal J(\params)= \frac{1}{M}\sum_{\{a_j^{\{k\}}\}_{j=1}^q \in B } \nabla_\params \bigg[ \mathcal G^d_k + \mathcal G^c_k + \beta^{-1}_{{S}} (\SD_k + \SC_k)\bigg]. \]
        \STATE Update weights $\params\leftarrow\params + \eta_t\nabla_{\params}\mathcal J(\params)$.
        \ENDFOR
        \ENDFOR
    \end{algorithmic}
    \vspace{0.5em}
\end{algorithm}

We also conducted coarse hyperparameter sweeps to find the optimal values for the hyperparameters of \agentname{}, cf.~Table~\ref{tab:model_hp}. We use a batch size of $128$ to train the policy. The policy network is optimized using Adam. The initial learning rate is set to \num{5e-4}, which is typical when training autoregressive networks; we employ a learning rate decay schedule which decreases by $98\%$ every $50$ iterations. The Autoregressive network is implemented using uniform masks and dense layers~\cite{germain2015made}.
The base layer (see Fig.~\ref{fig:arnet}) consists of two hidden layers with 100 neurons each and the heads contain $3\vert \Ad  \vert$ neurons in total.

The agent is trained via proximal policy gradient (PPO). We use four PPO updates to the policy network parameters per iterations. The clipping parameters are set as $\epc=0.1$ for the continuous policy, and $\epd=10^{-3}$ for the discrete policy. We include entropy bonus to increase exploration; the corresponding temperature schedule $\beta^{-1}_{\mathcal S}$ starts at \num{1e-1}, and drops by $99\%$ every $50$ iterations.

\begin{table}[h]
    \caption{\agentname{} Hyperparameters.}
    \label{table:ppo_params}
    \vskip 0.15in
    \begin{center}
        \begin{small}
            \begin{sc}
                \begin{tabular}{rl}
                    \toprule
                    Hyperparameter                                      & Value                      \\
                    \midrule
                    Optimizer                                           & Adam~\cite{kingma2014adam} \\
                    Learning rate                                       & \num{5e-4}                 \\
                    Likelihood ratio clip, $\epsilon^\nu$               & 0.1 ($\epc$)               \\
                                                                        & 0.001 ($\epd$)             \\
                    PPO Epochs                                          & 4                          \\
                    Hidden units (masked dense layer)                   & $[100, 100]$               \\
                    Activation function                                 & ReLU                       \\
                    baseline exponential moving average ($m$)           & 0.95                       \\
                    Learning rate annealing steps                       & 50                         \\
                    Learning rate annealing factor                      & 0.98                       \\
                    Learning rate annealing style                       & Staircase                  \\
                    Entropy bonus temperature ($\beta_{S, \{0\}}^{-1}$) & $1 \times 10^{-1}$         \\
                    Entropy bonus temperature decay steps               & 50                         \\
                    Entropy bonus temperature decay factor              & 0.99                       \\
                    Entropy bonus temperature decay style               & Smooth                     \\
                    Minibatch size                                      & 128                        \\

                    \bottomrule
                \end{tabular}
            \end{sc}
        \end{small}
    \end{center}
    \vskip -0.1in
    \label{tab:model_hp}
\end{table}

\begin{figure}[t!]
    \centering
    \centerline{
        \includegraphics[width=1.1\textwidth]{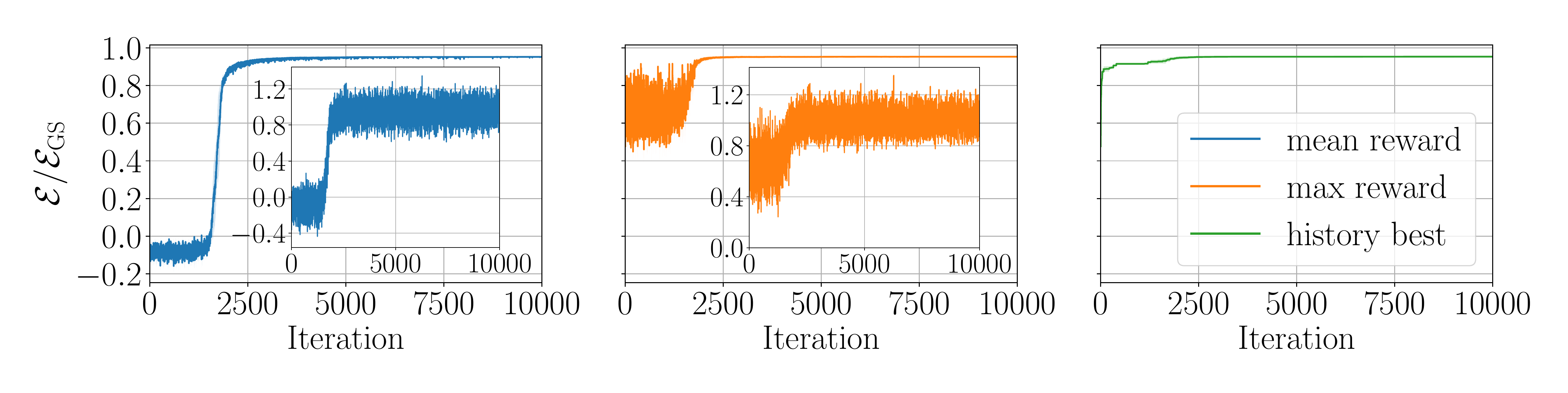}}
    \caption{ Spin-$1/2$ Ising model:
        training curves for \agentname{} with energy minimization as a cost function. The quantities in the main figure are noiseless evaluation, while in the inset are noisy measurement. The noiseless quantities are only for the evaluation's purpose, and the agent can only access the noisy quanties~(in the inset). The mean reward (blue curve) is the average energy ratio across the minibatch sampled from the autoregressive policy; the max reward (orange curve) is taking the maximum across the minibatch; the history best bookkeeps the best ever max reward during the training.
        The total duration is $T\!=\!10$ and the number of spin-$1/2$ particles is $N\!=\!8$. The discrete \agentname{} action space is $\Ad=\{H_1, H_2; Y,X|Y,Y|Z \}$, and we use $q\!=\!8$. Here, the noise is classic gaussian noise, with the noise level $\gamma\eq0.1$. }
    \label{fig:nn-training}
\end{figure}

A typical learning curve in the noisy setting is shown in Fig.~\ref{fig:nn-training}. Three quantities are recorded to measure the performance of the agent. In the noise setting, these quantities correspond to the ideal noise-free case. We use them only for the purpose of evaluation; during the training, the RL agent only has access to the noisy rewards. These quantities are shown in terms of the energy ratio with respect to the target ground state so the possible maximum is upper bounded by one; since the energy of a state can be either positive or negative, while the GS has a negative value, negative ratios are possible. Figure~\ref{fig:nn-training} shows that the agent starts to pick up the learning signal around two thousand iterations. After that, it slightly modifies the policy in order to achieve a higher reward.
Here, the mean reward stands for the sample mean of energy density at every iteration; the max reward is the maximum over the sample; the history best is the best-encountered reward during the entire training process.

\section{A comparison of compactly supported distributions defining continuous actions} \label{app:sigmoid-gaussian}

\subsection{Sigmoid Gaussian Distribution}
\label{app:sigmoid-gaussian-1}

In order to enforce the bounds for the duration output from the distribution, we apply the sigmoid function. This kind of finite bound of the action does help a lot in practice when we then normalize the actions to have the finite sum; otherwise, we observe a large variance when we normalize the total protocol duration to $T$ (see main text). To this end, we apply the sigmoid function to the Gaussian distribution. In the following formula, we have $x = f(y)$, where $f(y) = \frac{1}{1 + e^{-y}}$ is the sigmoid. We denote the original distribution as $\pi_0(y; \kappa, \xi)$ and the distribution after the transformation, as $\pi(x; \kappa, \xi)$.
$$
    \pi(x; \kappa, \xi)=\pi_0(y; \kappa, \xi)\left|\operatorname{det}\left(\frac{\mathrm{d} x}{\mathrm{d} y}\right)\right|^{-1}
$$
For example, if we choose $\pi_0$ to be Gaussian distribution according to $\mathcal N(\kappa, \xi^2)$, then
\begin{equation}
    \log \pi(x; \kappa, \xi)=-\log \xi-\frac{1}{2} \log (2 \pi)-\frac{1}{2}\left(\frac{\logit(x)-\kappa}{\xi}\right)^{2}- \log (x \left(1-x\right))
\end{equation}
Here, the logit function, $\logit(x) \eq \log x - \log(1-x)$, is the inverse of the sigmoid function $f(x) = 1/(1+\exp(-x))$.

Thus, the derivative with respect to the parameters (i.e.~$\kappa$ and $\xi$) can be computed analytically, and reads
\begin{align}
    \frac{\partial \log \pi(x ; \kappa, \xi)}{\partial \kappa} & =\frac{\logit(x)-\kappa}{\xi^2},                                       \\
    \frac{\partial \log \pi(x ; \kappa, \xi)}{\partial \xi}    & =-\frac1\xi+\frac{1}{\xi}\left(\frac{\logit(x)-\kappa}{\xi}\right)^{2}
\end{align}
We use this log probability in the policy gradient formula.

\subsection{Beta Distribution}
\label{app:beta}

The Beta distribution's probability density function is defined as:
$$
    \pi(x ; \kappa, \xi)=\frac{\Gamma(\kappa+\xi)}{\Gamma(\kappa) \Gamma(\xi)} x^{\kappa-1}(1-x)^{\xi-1},
$$
where the Gamma function is $\Gamma(z) = \int_0^\infty x^{z-1} e^{-t}\, dt$. Here, the $\kappa$ and $\xi$ are the parameters of the Beta distribution, which can be learned by the autoregressive policy network. The corresponding log-probability reads as
\begin{equation}
    \log \pi(x ; \kappa, \xi)=\log \Gamma(\kappa+\xi) - \log \Gamma(\kappa) - \log \Gamma(\xi) + (\kappa-1) \log(x) + (\xi-1)\log(1-x),
\end{equation}

Thus, the derivative with respect to the parameters (i.e.~$\kappa$ and $\xi$) reads
\begin{align}
    \frac{\partial \log \pi(x ; \kappa, \xi)}{\partial \kappa} & =\psi(\kappa+\xi) - \psi(\kappa) + \log(x),      \\
    \frac{\partial \log \pi(x ; \kappa, \xi)}{\partial \xi}    & =\psi(\kappa+\xi) - \psi(\xi) + \log(1\minus x),
\end{align}
where the digamma function is defined as the logarithmic derivative of the gamma function:
\begin{equation}
    \psi(x)=\frac{d}{dx}\ln\big(\Gamma(x)\big)=\frac{\Gamma'(x)}{\Gamma(x)}.
\end{equation}
Hence, the gradient can be used to compute the policy gradient using analytical expressions.

\section{Choosing the protocol duration $T$}

\label{app:durations}

\begin{figure}[h!]
    \centering
    \hspace{4em}
    \begin{tikzpicture}
        \begin{axis}[%
                line width=1.3pt,
                hide axis,
                xmin=10,
                xmax=50,
                ymin=0,
                ymax=0.4,
                legend style={draw=none, legend columns=-1}
            ]
            \addlegendimage{tblue}
            \addlegendentry{\; CD-QAOA\quad\quad \quad}
            \addlegendimage{torange}
            \addlegendentry{\; QAOA\quad\quad \quad};
            \addlegendimage{tgreen}
            \addlegendentry{\; adiabatic};
        \end{axis}
    \end{tikzpicture}
    \\
    \vspace{-0.5em}
    \includegraphics[width=0.7\textwidth]{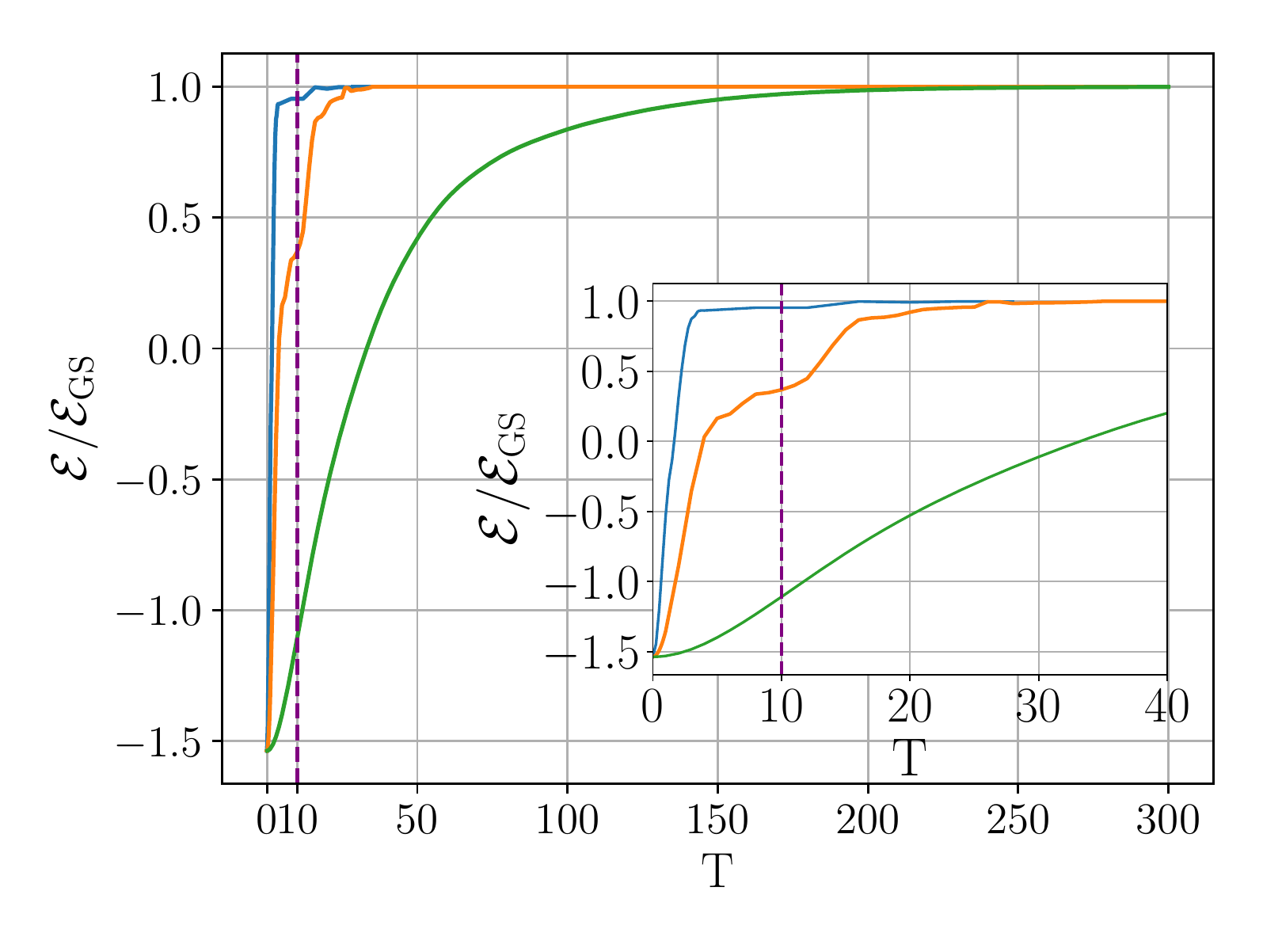}
    \caption{
        \small
        Spin-$1/2$ model: energy minimization at different protocol duration $T$ for three different methods in the nose-free setup: CD-QAOA (blue line), QAOA (red line), adiabatic evolution (green line). The physics model and the setting are the same as in  Sec.~\ref{sec:Ising-1/2}. For the adiabatic driving simulation, we used the protocol function  $\lambda(t)\!=\! \sin^2\left(\frac{\pi t}{2T}\right)$, $t\in[0,T]$. The quantum dynamics was solved for numerically, using a step size of $\Delta t\eq \num{1e-3}$. The system size is $N\eq4$.
        The vertical purple dashed line corresponds to $JT\eq10$.
    }
    \label{fig:scaling_T}
\end{figure}

Finally, let us explain the choice of protocol durations $JT=10$ used in our study. Figure~\ref{fig:scaling_T} shows a scan of the best energy over the protocol duration $T$ in the noise-free case for $N=4$ qubits for three methods: QAOA, CD-QAOA and adiabatic driving.
For the adiabatic driving, we consider the driven spin-$1/2$ Ising model:
\begin{equation}
    \label{eq:H_var_vs_qaoa}
    H(\lambda) \eq \lambda(t)H \! + \! ( 1-\lambda(t)) \tilde H,
\end{equation}
where $\lambda(t)\!=\! \sin^2\left(\frac{\pi t}{2T}\right)$ $t\in[0,T]$, is an smooth protocol satisfying the boundary conditions: $\lambda(0)\!=\! 0$, $\lambda(T)\!=\! 1$, $\dot\lambda(0)\!=\! 0 \!=\! \dot\lambda(T)$.
The initial state is the ground state at $t\!=\!0$, i.e.~$|\psi_i\rangle\!=\!|\!\!\uparrow\cdots\uparrow\rangle$, while the target state is the ground state of the Ising model at $t=T$ for $h_z/J=0.4523$ and $h_x/J=0.4045$. Here, $H$ is the target Hamiltonian defined in Eq.~\eqref{eq:IM}, and $\tilde H = -\sum_{i=1}^N S^z_i$.

The value $JT=10$ is selected to achieve a compromise: on one hand, it is large enough for CD-QAOA to reach close enough to the ground state; on the other hand, it is small enough for a discrepancy between the performance of CD-QAOA and QAOA to become clearly visible. Hence, $JT=10$ exemplifies nicely the benefits of using the generalized QAOA ansatz, as compared to QAOA. Last, we emphasize that both $JT=10$ are far away from the adiabatic regime, as shown by the adiabatic curve.

\end{document}